\documentclass[aps,pra,superscriptaddress,amsmath,amssymb,preprintnumbers,floatfix, showpacs,12pt]{revtex4}
\usepackage{amssymb}
\usepackage{epsfig}

\begin{document}
\title{ Non-local quantum field correlations and detection processes in QFT}
\author{Fabrizio Buscemi }
\email{fabrizio.buscemi@unimore.it}
\affiliation{ARCES, Alma Mater Studiorum, University of Bologna, Via Toffano 2/2, 40125 Bologna, Italy}
\affiliation{S3 Research Center, CNR-INFM, Via Campi 213/A, I-Modena 41100, Italy}
\author{Giuseppe Compagno }
\affiliation{Dipartimento di Scienze fisiche ed astronomiche
dell'Universit\`{a} di Palermo, Via Archirafi 36, 90123 Palermo,
Italy}
\begin{abstract}
Quantum detection processes in QFT must play a key role in the description of quantum 
field correlations, such as the appearance of entanglement, 
and of causal effects. We consider the detection 
in the case of a simple QFT model with a suitable interaction to exact treatment, consisting of a quantum scalar field
coupled linearly to  a classical scalar source. We then evaluate
the response function to the field quanta of two-level point-like quantum
model detectors,  and analyze the effects of the approximation adopted in standard detection
theory.  We show that the use of the RWA, that characterizes the Glauber detection model,
 leads in the detector response to non-local terms  corresponding to an instantaneously
spreading of source effects over the whole space. Other detector models, obtained with non-standard or the no-application of RWA, 
give instead local responses to field quanta, apart from
source independent vacuum contribution linked to preexisting correlations of zero-point field.

\end{abstract}
\date{\today}
 \maketitle


\section{Introduction}

The appearance of non local effects  in quantum mechanics
has received  great attention beginning from  the well-known 
EPR paradox \cite{Eins}. This indicates that the result of a measurement 
performed on one of a pair of correlated systems
has a non-local effect on the correlated physical measurement on the partner distant system.
Such a non-local behavior is related to the presence of quantum entanglement between the systems.
 Thus detection of    quantum correlations  between two  separated systems plays a key role in establishing 
 whether the systems are entangled or not. In particular models of measurements not causally  connected  are required to evaluate  a  genuine manifestation of the entanglement. In this case if  correlations are detected  they may violate Bell's inequality\cite{Bell} and therefore the two systems can be considered as entangled. Instead some models of measurement, leading by their same  nature to the instantaneous development  of non-local effects over the whole space, could give rise in their interpretation to the appearance of entanglement even in the absence of real  quantum correlations.

Another place where non-locality may manifest itself is in the
spacetime evolution of single particle wavefunction that gives
place to non zero contributions outside of the light-cone\cite{Heg3,Heg,Hege}. This
aspect of non-locality in quantum mechanics, with the building up of probability on space-like
distances, appears instead to give rise to a violation of causality. Non-local effects show up also  in Quantum Field Theory (QFT)   in the time evolution of initially localized quantum  field states both for  free fields and for interacting matter-field models \cite{Pri2,Heg3,Heg,Hege,Pri1,Bus3,Rubin,Buc}.

The use of appropriate  model of the detection process  for the interpretation
of the measurements in the observation  of non-local correlations plays a key role into evaluating the reality of these 
quantum correlations defining entanglement  or even of effects that appear to not satisfy the causal propagation of signals. 
In this context  it results thus to  be important the adoption of suitable 
quantum detectors models, and the appropriate detector model that must be adopted appears to
be still questioned \cite{milo,Mich,XTTT,XTTT2}.
Different   quantum  detector models  have been proposed in
literature. Among them the   Glauber  detector (GD) model, longly
used in photodetection theory and Quantum Optics \cite{Gla1}, and
the Unruh-DeWitt  detector (UDD) model utilized to describe
accelerated detectors and their excitations as response to the
inertial vacuum \cite{UNR,svai,sch,lag}. The GD model adopts the so-called
rotating wave approximation (RWA) whose application in the
solution  of QFT systems seems however to lead to the appearance of non-local effects \cite{Com1,Com3}.
In particular, for the case  of the interaction between atom and electromagnetic field within the dipole approximation, 
it has been shown that the use of the RWA
leads to the atomic dipole being coupled to the field at points other than the position of the dipole\cite{Cler}.  
From this point of view the  GD model could result to be unappropriate in models of matter-field interaction 
in describing the experimental observation aimed to detect quantum entanglement. In fact by its nature
this model gives  rise to the appearance 
of quantum correlations   over space-like distances which do not
represent a manifestation of a genuine entanglement. 
Thus the use of GD model also could lead to appearance of violation of the causal propagation of signals, even if 
the effective connection between RWA  and 
causality  in the Glauber detection theory is 
yet debated. In particular, it has been shown that 
 the photocounting probabilities for short
observation times appear to violate causality\cite{XTTT,XTTT2} and this    has led some authors  to suggest relevant modifications of
the Glauber photodetection theory \cite{XTTT,XTTT2}. Other investigations seem instead to
indicate that an appropriate use of the RWA in the GD model
guarantees causality \cite{milo}.

Thus the observability and the measurement of quantum correlations
in order to evidence entanglement and causal effects requires 
the use of appropriate detector models and in particular the adoption of a suitable Hamiltonian 
that not induces by itself non-locality.
The aim of this paper is to discuss the typical models adopted in describing 
quantum detection processes and their  relation to the possible appearance
of non-local effects in the context of QFT. To  this purpose in the first part of
the paper we  will analyze the GD model and the role played by the RWA into the appearance of effects over space-like distances.
Then, in order to connect the measurement of quantum correlations
to detection processes, we then shall analyze 
another suitable detector model and will obtain its response to the quantum field.
To this end here we shall consider  a system  consisting  of  a quantum
scalar field linearly interacting with a classical source
localized in a finite spacetime region \cite{kar,Mate,Bus,Bus2}.
Such a model, which can be exactly solved, appears to be of
interest because it allows us to have a clear view of the role played
by non-local effects in the quantum correlations  in the system 
without the limitations linked to the perturbative calculations.

The paper is organized as follows. In Sec.~\ref{model} we
illustrate the GD and UDD models, while in Sec.~\ref{model1}
a non standard application of the RWA to the  quantum detection of fields generated by
sources is  analyzed. In Sec.~\ref{model2} we shall introduce the
model of quantum scalar field coupled to a classical source and
then
 evaluate the response function of the GD
and UDD  to the field for different situations. Finally in
Sec.~\ref{model3} we comment the results obtained.

\section{Quantum detection models of scalar field}\label{model}

The quantum theory of photodetection, with the construction of a
model of detector, as
 developed  by Glauber \cite{Gla1}   has played   a key
role in Quantum Optics. However other kinds of detectors have also
been used in QFT, in particular,  by 
\cite{UNR,svai,sch,lag}. In both approaches the detectors are
particle detectors and the detection process represents  the
quantum measurement to detect the quanta of the field.   
Here    we shall utilize the GD and UDD  models    in the case of
scalar fields detection \cite{svai}.
\subsection{Unruh-De Witt scalar detector} UDD model\cite{UNR} is represented as an idealized particle of
negligible spatial extension and with   internal energy levels,
coupled  via a monopole interaction with a scalar field $\Phi(x)$.
The latter may be expressed  in terms of its positive and negative
frequency part as:
\begin{equation} \label{modena}
\Phi(x)=\Phi^{+}(x)+\Phi^{-}(x)
\end{equation}
where, taking $\hbar=1$ and $c=1$,
\begin{equation} \label{moden2}
\Phi^{+}(x)=\frac{1}{(2\pi)^{3/2}}\int
\frac{d^{3}\textbf{k}}{2\omega}\,a(\textbf{k}) e^{ -ik\cdot
x}\quad\textrm{and}\quad \Phi^{-}(x)={\Phi^{+}} ^{\dag}(x) .
\end{equation}
 $\omega= \sqrt{|\textbf{k}|^{2}+m^{2}}$  and   $a(\textbf{k})$,
$a^{\dag}(\textbf{k})$ are   respectively the usual annihilation
and creation operators that satisfy  the relativistic commutator
rules:
\begin{equation}
\left[ a(\textbf{k}),a^{\dag}(\textbf{k}')\right]=2\omega
\delta^{3}(\textbf{k}-\textbf{k}').
\end{equation}
The    detector is characterized by two energy levels $\omega_{g}$
and  $\omega_{e}$, with  eigenstates
 $|g\rangle$  and $|\,e\rangle$ respectively. It  moves
 along the line word line described by the function $x(\tau)$, with   $\tau$ the proper time.
The UDD model in the case of scalar fields   is defined  by the
 following interaction Hamiltonian:
\begin{equation}\label{interazio}
H^{int}_{UDD}=-c_{1}m(\tau)\Phi(x(\tau)),
\end{equation}
with $m(\tau)$ the detector monopole moment and
 $c_{1}$  the field-detector coupling constant. Notice that  $H^{int}_{UDD}$
 contains both conserving and non-conserving energy terms.
 We shall take the  interaction turned on only
for  a finite time interval   $\tau=\tau_f-\tau_i$ . The state of
the detector-field system at initial time $\tau_i$  is
 $|i\rangle= |g\,\psi_{i}\rangle=|g\rangle\otimes|\psi_i\rangle$
where  $|g\rangle$   is the  detector state ground and
$|\psi_i\rangle$ the field  state.

Using the interaction picture the first  order transition
amplitude from $|g\,\psi_{i}\rangle$ to  $|e\,\psi_{f}\rangle$
is:
\begin{equation} \label{ampro}
{A_{UDD}}_{|g\psi_{i}\rangle\rightarrow|e\psi_{f}\rangle}=\langle
e\,\psi_{f}|U(t)|g\,\psi_{i}\rangle=ic_{1}m_{eg}\int_{\tau_i}^{\tau_f}e^{i\omega_{eg}\tau
'} \langle \psi_{f}|\Phi(x(\tau '))|\psi_{i}\rangle d \tau '
\end{equation}
with $m_{eg}=\langle e|\widehat{m}(0)|g\rangle$ and
$\omega_{eg}=\omega_{e}-\omega_{g}$. Using the positive and
negative frequency parts of the field operator the matrix elements
 appearing within integral in Eq.~(\ref{ampro}) can be
written as:
\begin{equation} \label{ampro3}
  \langle
  \psi_{f}|\Phi^{+}(x(\tau '))|\psi_{i}\rangle+
  \langle \psi_{f}|\Phi^{-}(x(\tau '))|\psi_{i}\rangle ,
\end{equation}
with the first term describing   the absorption and the second
 the emission of field quanta by the detector. In the UDD both the terms  $\langle
  \psi_{f}|\Phi^{+}(x(\tau '))|\psi_{i}\rangle$ and  $\langle
  \psi_{f}|\Phi^{-}(x(\tau '))|\psi_{i}\rangle$ contribute to the detector excitation
 amplitude and correspond   respectively to detector excitation with absorption or emission of a field quantum. In   particular  the  second term
 represents the response of the
detector to the vacuum fluctuations. In order to have a
 better insight into the different kinds of processes  occurring in the
 scalar field-UDD interaction,  here we give the expression of
 the   amplitude probability of excitation UDD in terms of annihilation and creation operators
 of scalar quanta.  By inserting  Eq.~(\ref{moden2})
in  Eq.~(\ref{ampro}), we obtain:
\begin{eqnarray} \label{moden3}
\lefteqn{
{A_{UDD}}_{|g\psi_{i}\rangle\rightarrow|e\psi_{f}\rangle}= {}}\\
\nonumber
 & &\frac{ic_{1}m_{eg}}{(2\pi)^{3/2}}\int
\frac{d^{3}\textbf{k}}{2\omega}\int_{\tau_i}^{\tau_f} d\tau'
\left[ e^{ ik\cdot \textbf{x}(\tau)} e^{i(\omega_{eg}-\omega)\tau
'} \langle \psi_{f}|a(\textbf{k})|\psi_{i}\rangle + e^{ -ik\cdot
\textbf{x}(\tau)} e^{i(\omega_{eg}+\omega)\tau '} \langle
\psi_{f}|a^{\dag}(\textbf{k})|\psi_{i}\rangle \right],
\end{eqnarray}
where the emission of  quanta of the field with energy $\omega$,
is given in  the integrand by   the factor
$e^{i(\omega_{eg}+\omega)\tau '}$. The absorption process
instead leads
  to the factor  $e^{i(\omega_{eg}-\omega)\tau '} $ in the integrand of the  above expression.

The probability of detection of the UDD  is thus obtained  by
taking the square modulus of Eq.~(\ref{ampro}) and summing over
all the possible field final states:
\begin{equation}\label{afrikka}
P_{UDD}(\tau_f,\tau_i)=c^{2}_{1}|m_{eg}|^{2}\int_{\tau_i}^{\tau_f}\int_{\tau_i}^{\tau_f}d\tau
'd\tau '' e^{i\omega_{eg}(\tau ''-\tau ')} \langle
\psi_{i}|\Phi(x(\tau '))\Phi(x(\tau ''))|\psi_{i}\rangle
\end{equation}
From the above expression~  it comes out  that the response of
detector depends on the  motion of the detector itself, the well-known
Unruh effect is in fact related to this property.   The response  of   a
uniformly accelerated UDD with acceleration $\alpha$ to the vacuum
fluctuations is the same of a unaccelerated UDD
immersed in a bath of thermal radiation at temperature $T=1/(2\pi
k \alpha)$ \cite{UNR}.

\subsection{Glauber scalar detector}\label{modelGD}
The   GD model\cite{Gla1}, commonly  adopted  in quantum optics, is  obtained
by applying  the RWA in the interaction term. The  use of such an approximation, which
permits to easily evaluate the   photodetection probability,
is valid as long as the measurement time and pulse length  of
detected field are long compared to a typical optical cycle.

The RWA can analogously be applied for the case of scalar
detection in the field-detector interaction
Hamiltonian of Eq.~(\ref{interazio}). It reduces to the Hamiltonian
\begin{eqnarray}\label{interazio2}
H^{int}_{GD}=-c_{1}\left[m_{ge}(\tau)|g \rangle \langle e |\Phi^{-}(x(\tau))+m_{eg}(\tau)|e \rangle \langle g | \Phi^{+}(x(\tau))\right]
\end{eqnarray}
where the closure relation for detector eigenstates  $|g\rangle \langle g|+|e\rangle \langle e|= \mathbb{I}$ has been used and  $m_{e(g)g(e)}(\tau)=\langle e(g)|\widehat{m}(\tau)|g(e)\rangle$. Note that in the above expression  do not appear the  anti-resonant terms  $ m_{eg}(\tau)\, |e \rangle \langle g | \Phi^{-}(x(\tau))$  and $ m_{ge}(\tau)\, |g \rangle \langle e | \Phi^{+}(x(\tau))$,      which describe the creation  of scalar quanta with excitation of the detector and the annihilation  of scalar quanta with the decay of the detector, respectively. In fact
the  RWA implies the neglection of  such  counter-rotating terms.

According to this model the detection is only considered as  an
absorption process. As  seen from Eq.~(\ref{moden3}), in the term
describing   the emission of  quanta of the field   the factor
$e^{i(\omega_{eg}+\omega)\tau '}$ appears, which is rapidly
oscillating
 and  gives  a negligible contribution for  $(\tau_f-\tau_i) \gg 1/w_{eg} $. In this sense
such a   process can be considered   virtual, since it can occur
only for short time intervals $(\tau_f-\tau_i)$ obeying
$\omega_{eg} (\tau_f-\tau_i)\lesssim 1$ and moreover does not
conserve energy. Instead  the  absorption process is given in  the integrand by   the factor  $e^{i(\omega_{eg}-\omega)\tau '} $. The 
adoption of RWA  forbids
the virtual transitions where the energy is
not conserved. This  implies that the  only term,  that  now comes out in
Eq.~(\ref{ampro}), is the matrix element $ m_{eg}\, \langle
\psi_{f}|\Phi^{+}(x(\tau '))|\psi_{i}\rangle$, thus only the
positive frequency part of the field appears in the  first order amplitude transition from the initial state   $|g\rangle|\psi_{i}\rangle$ to
 $|e\rangle|\psi_{f}\rangle$ for the   GD  model. This  is given by
 \begin{equation} \label{ampro2}
A_{GD_{|g\psi_{f}\rangle\rightarrow|e\psi_{i}\rangle}}
=ic_{1}m_{eg}\int_{\tau_i}^{\tau_f}e^{i\omega_{eg}\tau '} \langle
\psi_{f}|\Phi^{+}(x(\tau '))|\psi_{i}\rangle d \tau '
\end{equation}
and   then leads to the  probability detection:
\begin{equation}\label{ense}
  P_{GD}(\tau_f,\tau_i)=c^{2}_{1}|m_{eg}|^{2}\int_{\tau_i}^{\tau_f}\int_{\tau_i}^{\tau_f} d \tau ' d \tau ''e^{i\omega_{eg}(\tau''- \tau')}
\langle
\psi_{i}|\Phi^{-}(x(\tau '))\Phi^{+}(x(\tau ''))|\psi_{i}\rangle .
\end{equation}
The response of the GD to the vacuum field state $| 0 \rangle$  is
\begin{equation}\label{ensel}
  P_{GD}(\tau_f,\tau_i)=c^{2}_{1}|m_{eg}|^{2}\int_{\tau_i}^{\tau_f}\int_{\tau_i}^{\tau_f} d \tau ' d \tau ''e^{i\omega_{eg}(\tau''- \tau')}
\langle
 0|\Phi^{-}(x(\tau '))\Phi^{+}(x(\tau ''))|0\rangle =0
\end{equation}
The GD, as a consequence of the RWA,  does therefore not feel the
zero point vacuum fluctuations. From Eq.~(\ref{ensel}) it follows also
that the  response of the GD to the vacuum    does not  depend
from the state of the motion  of the same detector. The detection probability  vanishes in
particular for detectors travelling either along inertial  or
accelerated world lines.  Therefore such a detector can not show
the well known Unruh effect.

The adoption  of the GD model, with its use of the RWA, to detect quantum correlations due to entanglement 
appears to be controversial in QFT systems\cite{XTTT,XTTT2,milo}. By applying an approach
 already used  in the photodetection of free electromagnetic fields \cite{XTTT,XTTT2},  here we want to show  the appearance of non- local effects in the  Glauber scalar   detection theory. To this aim  we examine a free quantum scalar field  $| \Psi \rangle $,  expressed as
 \begin{equation} \label{musik}
| \Psi \rangle =\int
\frac{d^{3}\textbf{k}'}{2\omega}\, |\alpha(\textbf{k}')\rangle
\end{equation}
where   the state $ |\alpha(\textbf{k}')\rangle$
is  the    eigenstate  of the annihilation operator
$a(\textbf{k})$ with eigenvalue $\alpha(\textbf{k})$
\begin{equation}\label{musik2}
a(\textbf{k})|\alpha(\textbf{k}')\rangle=\alpha(\textbf{k})|\alpha(\textbf{k}')\rangle.
\end{equation}
and therefore $| \Psi \rangle $ is the coherent state satisfying $a(\textbf{k})| \Psi \rangle=\alpha(\textbf{k})| \Psi \rangle$.
Taking into account   Eqs.~(\ref{modena}), (\ref{musik}) and (\ref{musik2}), the action of the operator $\Phi^{+}(x)$ on $| \Psi \rangle $ gives
\begin{equation} \label{moden5}
\Phi^{+}(x)| \Psi \rangle = V(x) | \Psi \rangle ,
\end{equation}
where
\begin{equation} \label{moden6}
V(x)=\frac{1}{(2\pi)^{3/2}}\int \frac{d^{3}\textbf{k}}{2\omega}\,
e^{ i(\textbf{k}\cdot \textbf{x} -\omega t)}\alpha (\textbf{k}).
\end{equation}
It may easily be shown that $V(x)$ satisfies the classical homogeneous Klein-Gordon
equation
\begin{equation}
\left(\frac{\partial}{\partial t^2}  -\frac{\partial}{\partial
\textbf{x}^2}  + m^2\right)V(x)=0,
\end{equation}
 and  can
be therefore interpreted as a classical signal  propagating freely.
In Eq.~(\ref{moden6})  it appears only the factor  $e^{ -i\omega t}$
with $\omega > 0$.   Extending $t$ to a complex variable,this corresponds in the complex   plane
$t=t_{1}-it_{2}$, to the appearance of the term   $e^{ -\omega t_{2}}$.
    $V(x)$ is thus an analytical function in the lower complex
$t$ halfplane  and then its real and imaginary parts are therefore related
by a Hilbert transformation,
\begin{equation} \label{moden7}
\textrm{Im} \,V(\textbf{x},t)=\frac{1}{\pi}\int_{-\infty}^{\infty}
\frac{\textrm{Re}  \,V(\textbf{x},t')}{t-t'} dt' .
\end{equation}

Now we can  easily evaluate  the time evolution of quantities of interest in terms of  $V(x)$.
The first order Glauber correlation function  is defined as
\begin{equation} \label{moden8}
G(x,x)= \langle \Phi^{-}(x)\Phi^{+}(x) \rangle ,
\end{equation}
and therefore is given by the expectation value of the operator $\Phi^{-}(x)\Phi^{+}(x) $ on the quantum field state.  Such  a function permits  to estimate the rate of scalar quanta counting probability. By inserting Eqs.~(\ref{musik}) and (\ref{musik2}) in (\ref{moden8}), we obtain for it:
\begin{equation} \label{stage2}
G(x,x)=|V(x)|^{2}=(\textrm{Re}\,V(x))^{2}+(\textrm{Im}
\,V(x))^{2}.
\end{equation}
The mean value of the scalar field operator $\Phi(x)$ has instead the form:
\begin{equation} \label{stage}
\langle \Psi|\Phi(x)| \Psi \rangle = 2 \,\textrm{Re} \,V(x) .
\end{equation}

Let us consider a signal $V(x)$ consisting of a plane wave  with a sharp
front moving  in the positive $\bf{z}$ axis direction whose the real part is given by
\begin{equation}\label{stage3}
\textrm{Re} V(x)= \Theta (t-\textbf{z}) f(\textbf{z}-t),
\end{equation}
 where $\Theta$ is the Heaviside function and 
 \begin{equation}
 f(\textbf{z}-t)=\left .\begin{array}{c} f_0 
 \quad\textrm{for} \quad t-\textbf{z}\in [0,\Delta \bar{\textbf{z}}]\\
0\quad\textrm{for} \quad t-\textbf{z}\notin [0,\Delta \bar{\textbf{z}}] \end{array} \right\}
\end{equation}
with $\Delta \bar{z}$ indicating the  signal length here assumed to be small. We note from Eqs.~(\ref{stage}) and ~(\ref{stage3}) that 
$\textrm{Re} V $ and therefore the mean value of the field reaches the detector  at time $t=\bf{z}$ and is equal
to 0 for $t<\bf{z}$.  $\textrm{Im} V$, related by Eq.~(\ref{moden7}) to $\textrm{Re} V $ will be given by
\begin{equation}
\textrm{Im} V(x)= \frac{f_0}{\pi}\ln{\left|\frac{t-\textbf{z}}{t-\textbf{z}-\Delta \bar{\textbf{z}}}\right|}.
\end{equation}
It results to  differ from zero for all $t$ even if $\textrm{Re} V =  0$ for $t<\bf{z}$.  Therefore
the Glauber correlation function $\langle \Phi^{-}(x)\Phi^{+}(x) \rangle$  written in Eq.~(\ref{stage2}) does not vanishes before  $\langle\Phi(x)\rangle$
for $t<\bf{z}$.  Such a
result implies that the GD model, leading by its  very
nature to the development of effects over space-like distances, is unappropriate to detect both  the appearance of entanglement and causality 
 in the time evolution of  free fields
\cite{XTTT,XTTT2}.

\section{A non standard application of the RWA to
  quantum detection  theory}\label{model1}
In the previous section we have seen that non-locality  shows
up at level of   Glauber detection of free scalar fields. This
induces   first to inquire  if such a behavior may be observed
when   other kinds of detectors are
used, and then  to examine the detection
processes in the case of field generated by  quantum sources.

In particular,  we shall here investigate the  role
played  by  a  ``non standard''
application   of the RWA  to the quantum
detection theory for  the case of quantum scalar fields
interacting with sources. Starting from   a complete Hamiltonian
$H$, which contains conserving  and non-conserving energy terms,
the detection probability rate for the UDD point-like at rest and
localized at $\textbf{x}$ with $x(\tau)=x=(\textbf{x},t)$
 is given by the time derivative of Eq.~(\ref{afrikka}) and can be expressed as
\begin{equation} \label{mi2}
  \dot{P}_{UDD}(t) = 2 c_1^2 |m_{eg}|^{2}\textrm{Re}\int_{0}^{t} d t'\langle
  \psi_{i}|\Phi(\textbf{x},t)\Phi(\textbf{x},t')|\,
  \psi_{i}\rangle e^{i\omega_{eg}(t'-t)}
\end{equation}
where we have assumed that the field detector interaction is
turned from $\tau_i=0$ to $\tau_f=t$.  We shall analyze the detection probability
rate in the Heinseberg picture. It has  again  the form of Eq.~(\ref{mi2}) where  $\Phi'(\textbf{x},t)$ is now the
Heisenberg operator satisfying the equation of the motion:
\begin{equation} \label{mi3}
\frac{\partial\Phi'\textbf{x},t)}{\partial t}=i
[H,\Phi'(\textbf{x},t)]
\end{equation}
Following the same approach previously adopted in
QED \cite{milo}, a  formal solution of  Eq.~(\ref{mi3}) can be
expressed  by writing  the Heinseberg operator
$\Phi'(\textbf{x},t)$ as
\begin{equation}
\Phi'(\textbf{x},t)=\Phi'_{0}(\textbf{x},t)+\Phi'_{RR}(\textbf{x},t)+\Phi'_{s}(\textbf{x},t)
\end{equation}
with   $\Phi'_{0}(\textbf{x},t)$  the free field,
$\Phi'_{RR}(\textbf{x},t)$  the radiation reaction field of the
detector on itself while $\Phi'_{s}(\textbf{x},t)$ indicates the
field due to the source. The retarded source-field  can be
expressed as
$\Phi'_{s}(\textbf{x},t)=F'(\textbf{x},t)\Theta(t-r)$, where $\Theta$ is the Heaveside function
guaranteeing causality and
therefore
\begin{equation}\label{campioni}
\Phi'(\textbf{x},t)=\Phi'_{0}(\textbf{x},t)+\Phi'_{RR}(\textbf{x},t)+F'(\textbf{x},t)\Theta(t-r).
\end{equation}
In the above expression  we have assumed  the external field
source to be at distance $r$ from  the point-like detector
localized at $\textbf{x}$.

Now let us define
\begin{equation} \label{allevi1}
\widetilde{\Phi'}^{+(-)}(\textbf{x},t)=\Phi'^{+(-)}_{0}
(\textbf{x},t)+\Phi'^{+(-)}_{RR}(\textbf{x},t)+\widetilde{\Phi'}^{+(-)}_{s}(\textbf{x},t),
\end{equation}
where $\Phi'^{+(-)}_{0} $ and $\Phi'^{+(-)}_{RR} $ are
 the positive (negative) frequency parts of the  free and reaction radiation field
 respectively and $\widetilde{\Phi'}^{+(-)}_{s}$ is
\begin{equation} \label{allevi2}
\widetilde{\Phi'}^{+(-)}_{s}(\textbf{x},t)=F'^{+(-)}(\textbf{x},t)\Theta(t-r)
\end{equation}
with $F'^{+(-)}$ indicating the positive (negative) frequency part
of $F'$.

For time intervals larger than $1/\omega_{eg}$ we can adopt the
approximation  already used by Milonni \emph{et al.}  to  treat
the electromagnetic field case \cite{milo}.  This consists in
approximating   Eq.~(\ref{mi2})   with the expression
\begin{equation}\label{milonni2}
\dot{P}_{UDD}(t)\simeq\dot{P}_{MD}(t)= 2 |m_{eg}|^{2}
\textrm{Re}\int_{0}^{t} d t'\langle
  \psi_{i}|\widetilde{\Phi'}^{-}(\textbf{x},t)\widetilde{\Phi'}^{+}(\textbf{x},t)|  \psi_{i}\rangle
  e^{i\omega_{eg}(t'-t)}\quad\text{for}\quad t\gg 1/\omega_{eg},
\end{equation}
which can be considered as   the rate detection probability,
evaluated in the Heinseberg picture, of  a new scalar quantum
detector model, that is the  ``Milonni detector'' (MD). In
Eq.~(\ref{milonni2}), instead of the field operator
$\Phi(\textbf{x},t)$ which appears in  Eq.~(\ref{mi2}) and thus 
contains also terms including positive and negative frequency parts ,
such as $\Phi'^{+}\Phi'^{-}$, only  the combination $\widetilde{\Phi'}^{-}\widetilde{\Phi'}^{+}$ is present.

We stress that  the approximation  used in Eq.~(\ref{milonni2}),
has been applied only after calculating the fields based on full
Hamiltonian including  conserving and non-conserving energy terms.
Now we will show that this way of using such an approximation  represents a non
standard application of the RWA as originally performed  in
Glauber formulation and as a matter of fact a different one. In fact from the Eq.~(\ref{allevi2}) we
observe that while $\widetilde{\Phi'}^{+(-)}_{s}$ gives the retarded
positive (negative) frequency part of the external source field
it does not coincide with positive (negative) frequency part of
the retarded operator $\Phi'_{s}$ which should be inserted
according to  the standard application of the RWA. Indeed   the $\Theta$
function by itself consists of positive and negative frequency
parts as:
\begin{equation}
\Theta(\tau)=\Theta^- (\tau)
+\Theta^+(\tau)=\lim_{\epsilon\rightarrow 0}\frac{-1}{2\pi
i}\left\{\int_{-\infty}^{0}\frac{d\omega\,e^{-i\omega\tau}}{\omega+i\epsilon}
+\int_{0}^{\infty}\frac{d\omega\,e^{-i\omega\tau}}{\omega+i\epsilon}\right\}.
\end{equation}
Therefore the causally retarded source field $\widetilde{\Phi'}^{+(-)}_{s}$
contains both positive and negative frequency components. The   approach  here  described that 
replaces in the detection probability rate of Eq.~(\ref{milonni2})
the full retarded fields with the retarded positive (negative)
frequency part of the field is
 different  from  the standard form of the RWA 
 that is performed at the beginning in the Glauber detection theory
 in the Hamiltonian of the
system. This gives place to the detection rate probability of the defined MD, whose use 
prevents the development of quantum correlations over space-like distances
as we now will show.

By taking in     the rate detection probability of MD, given by
Eq.~(\ref{milonni2}), as the initial field state  the vacuum state
$|\psi_{i}\rangle=|\,0\rangle$  and then  inserting
Eqs.~(\ref{allevi1}) and  (\ref{allevi2}) we obtain
\begin{eqnarray}\label{mileonni3}
\lefteqn{\dot{P}_{MD}(t)= 2 |m_{eg}|^{2}\textrm{Re}\int_{0}^{t} d
t' \Bigg[\langle
\Phi'^{-}_{RR}(\textbf{x},t)\Phi'^{+}_{RR}(\textbf{x},t')\rangle
{} }\nonumber \\
& & {} + \Theta(t'-r) \langle
\Phi'^{-}_{RR}(\textbf{x},t)F'^{+}(\textbf{x},t')\rangle +
\Theta(t-r) \langle
F'^{-}(\textbf{x},t)\Phi'^{+}_{RR}(\textbf{x},t')\rangle \nonumber
\\ & & {} +\Theta(t-r)\Theta(t'-r)\langle
F'_{-}(\textbf{x},t)F'^{+}(\textbf{x},t')\rangle\Bigg]
e^{i\omega_{eg}(t'-t)}
 \end{eqnarray}
Under the assumption that the monopole detector atom is
only weakly perturbed  the above expression becomes, similarly to
the electromagnetic field case \cite{milo}:
\begin{equation} \label{mozart}
\dot{P}_{MD}(t) \cong  2
|m_{eg}|^{2}\Theta(t-r)\textrm{Re}\int_{r}^{t} d t'\langle
F'^{-}(\textbf{x},t)F'^{+}(\textbf{x},t')\rangle
e^{i\omega_{eg}(t'-t)}
\end{equation}
The presence of the function $\Theta(t-r)$ in the rate probability
expression  guarantees that the influence of the source-field is
is not vanishing only inside the light-cone centered on the external field source.
 Thus  the adoption of the MD for models of matter-field
interaction being the sources quantum, like the in Fermi model
\cite{Pow,Pri1}, or classical like in other models \cite{kar,Mate,Bus}
does not lead to quantum correlations spreading in the whole space and is moreover
 causal , even if such a behavior  appears to be ``forced'' by the
 approximation used in  rate detection probability.

\section{The quantum detection in a scalar QFT model}\label{model2}
An analysis of  the
measurement and the possible observability of quantum
correlations must use  suitable detectors and can be strictly accomplished
within  exactly solvable physical models. In this spirit  a simple QFT system, consisting of a quantum scalar
field coupled to a classical source,  has been recently
investigated with none of the limitations related to  perturbative
calculations. So
  it appears  of interest to
study for this system    the response of the  various  detectors to
the field generated by localized sources.
\subsection{The model }
We consider  a QFT model of a   quantum scalar field $\Phi(x)$
linearly interacting with a classical scalar source $j(x)$,
assumed to be localized
 in a finite  spacetime region and turned for a finite time\cite{kar,Mate,Bus}.
  The Hamiltonian term  describing the interaction is given by:
\begin{equation}\label{oxa}
H_{int}(t)  =  g\int
_{-\infty}^{+\infty}d^{3}\textbf{x}\bigg(\Phi^{+}(\textbf{x},t)+\Phi^{-}
(\textbf{x},t)\bigg)j(\textbf{x},t)=H_{int}^{+}(t)+H_{int}^{-}(t)
\end{equation}
where $g$  is the source-field coupling constant.

Initially ($t=0$) the field is taken in its vacuum state
$|0\rangle$. The state $|t\rangle$, describing the  system at time
$t$, will be
\begin{equation} \label{opevu}
|t\rangle=U(t)|0\rangle
\end{equation}
where $U(t)$ is the interaction picture time evolution operator.
Solving the equation of motion that derives  from  Eq.~(\ref{oxa})
we   get  for $U(t)$ a formal expression valid at all orders in
$g$ as:
\begin{equation}\label{aiduc}
U(t)=\exp\left(-i\int_{0}^{t} d t'
H_{int}^{-}(t')\right)\exp\left(-i\int_{0}^{t} d t'
H_{int}^{+}(t')\right)e^{-\xi(t)}e^{\alpha(t)} .
\end{equation}
In Eq.~(\ref{aiduc}) the coefficients  $\alpha(t)$ ,  $\xi(t)$
depend explicitly on the source as:
\begin{eqnarray}\label{stefrub}
\alpha(t) & =&
\frac{ig^{2}}{2}\int_{0}^{t}dt_{1}\int_{0}^{t}dt_{2}\int
d^{3}\textbf{x}_{1} \int \!\!
    d^{3}\textbf{x}_{2}\,j(\textbf{x}_{1},t_{1})\Delta_{-}(\textbf{x}_{1}-\textbf{x}_{2},t_{1}-t_{2})j(\textbf{x}_{2},t_{2})\\
\xi(t)&=&\frac{ig^{2}}{2}\int_{0}^{t}\!\!dt_{1}\int_{0}^{t}\!\!dt_{2}\nonumber
 \int\!\!d^{3}\textbf{x}_{1}\int\!\!
d^{3}\textbf{x}_{2}\,\Delta(\textbf{x}_{1}-\textbf{x}_{2},t_{1}-t_{2})j(\textbf{x}_{1},t_{1})j(\textbf{x}_{2},t_{2})\Theta(t_{1}-t_{2})
\end{eqnarray}
where $\Delta$ is the two-point function, given by the field
commutator as $[\Phi(x),\Phi(y)]=i\Delta(x-y)$,
  and $\Delta_{-}$ is its negative frequency part \cite{Bjo2,Pes}.

It has been previously shown that the dynamics of   any local observable
$\widehat{O}(\Phi(x),
\partial_{\mu}\Phi(x))$, satisfying the micro-causality principle
and represented by an analytical function of the field operator
and its space and time derivatives, depends  causally on the
source\cite{Bus2,Bus}. 
With this model the presence of non-locality has also been  investigated by analyzing the
localization properties of average values of local operators
in connection to  Hegerfeldt's theorem
\cite{Heg3,Heg,Hege} which seems to imply causality violation  for 
the time evolution of the wavefunctions, and one-point positive
localization observables. In the same spirit and in the connection
to the relevance of the detection theory for relating
the results of measurements with the form of quantum correlation functions
here we will evaluate
the expectation values, on the quantum state $|t\rangle$
describing the  system, of the Glauber and Newton-Wigner
operators, which have recently been used in QFT models  both of
free fields and of matter-field interaction \cite{Pri1,Pri2}.

The  Glauber operator for the scalar field is defined as
$\widehat{\rho}_{G}(x)=  \Phi^{-}(x)\Phi^{+}(x)$ and its
expectation value   on the state  $|t\rangle$ is
\begin{eqnarray} \label{bisga}
\langle t |\widehat{\rho}_{G}(x)|t\rangle=g^{2}\widetilde{\Delta}_{+}(x-y)\widetilde{\Delta}_{-} (x-y)
\end{eqnarray}
where the function
$\widetilde{\Delta}_{+}\left(\widetilde{\Delta}_{-}\right)$,
defined as
\begin{equation}\label{zeras}
\widetilde{\Delta}_{\pm}(x-y)\equiv\int_{0}^{t} d t'\int
d^{3}\textbf{x}'\Delta_{\pm}(\textbf{x}-\textbf{x}',t-t')j(\textbf{x}',t'),
\end{equation}
 is not zero outside the  light cone containing the source. Therefore  the expectation value
 of $\widehat{\rho}_{G}$ given by   Eq.~(\ref{bisga}) does not
 show a causal behavior.

The Newton-Wigner operator for scalar field has instead the form
\cite{newi, Pri1,Fle}:
\begin{equation}
\rho_{NW}(x)=a^{\dag}_{NW}(x)a_{NW}(x)
\end{equation}
where  $a^{\dag}_{NW}(x)$ and  $a_{NW}(x)$  may be expressed in terms
of the negative(positive) frequency part of the field operator
$\Phi(x)$ as
\begin{eqnarray}
a^{\dag}_{NW}(x)&=&R(\textbf{x})\Phi^{-}(x)\nonumber\\
a_{NW}(x)&=& R(\textbf{x})\Phi^{+}(x)
\end{eqnarray}
where
\begin{equation}\label{Kompat}
R(\textbf{x})=\sqrt{2}\left(m^{2}-\left(\frac{\partial}{\partial\textbf{x}}\right)^{2}\right)^{1/4}.
\end{equation}
$R(\textbf{x})$ is a non local operator  that  may be shown to
correspond to a non local integral transformation\cite{Fle}. The
expectation value of  $\rho_{NW}(x)$ on $|t\rangle$ is:
\begin{eqnarray}\label{maumos3}
   \langle t |\rho_{NW}(x)|t\rangle =g^{2}R(\textbf{x})
\widetilde{\Delta}_{+}(x-y)R(\textbf{x})\widetilde{\Delta}_{-}
(x-y)
\end{eqnarray}
The expectation value of the Newton Wigner operator $\rho_{NW}$ on
$|t\rangle$ immediately shows a local behavior.  In fact  it
contains the action of the non local operator $R(\textbf{x})$ on
the functions $\widetilde{\Delta}_{+}(x-y)$ and
$\widetilde{\Delta}_{-}(x-y)$ which already, present contributions
outside the light-cone centered on  the source. Non-local effects
shown by  both   the Glauber and Newton-Wigner operators are 
however attributable to the fact  that these operators do not
satisfy the micro-causality principle \cite{Bus}. This implies that
the  measurement on one spacetime point has influence  another
point at a space-like distance.

\subsection{Response of UD detector}
The appearance in the scalar  model of non-local effects seems    to
be at variance with the  results found  in previous works
that use  local operator functions of the field and of its time
and space derivatives \cite{Bus2,Bus}. However they are connected
to the use of localization operators that do not satisfy the
micro-causality principle. All of this stresses once more the key
role played by a proper detection theory in the question
concerning non-locality and  measurement  of  quantum correlations due to
to a genuine entanglement and of causal effects.
Here we will calculate explicitly the response of the  point-like
UD detector to the field in our QFT scalar model.

In order to keep  the problem simple we will assume  the detector
at rest  at  space point $\textbf{x}$, so that the function
describing its  world line becomes $x(\tau)=x=(\textbf{x},t)$.
Therefore  the effects,  that depend from  the motion of the
detector, as the Unruh ones,  will not appear in the detection
probability. Moreover we shall assume that the  source
 coupled to the quantum field is classical and localized within  a  sufficiently small spacetime region
 around the
spacetime point $y=(\textbf{y},y_{0})$. This source is thus effectively point-like and we shall assume that it is turned  on and off for
an infinitesimal time interval. In this case  from
Eq.~(\ref{aiduc}), the quantum field state  describing  the
evolving system at  time $t$ takes the form:
\begin{equation}\label{sean}
|t \rangle =\exp{\Big(-ig\Theta(t-y_{0})\Phi^{-}(y)\Big)}|0\rangle
e^{\alpha_{0}(t)}
\end{equation}
with
\begin{equation} \label{sean2}
\alpha_{0}(t)=\frac{ig^{2}}{2}\,\Theta^{2}(t-y_{0})\lim_{x\rightarrow
0 }\Delta_{-}(x) .
\end{equation}
The above expression for $\alpha_{0}(t)$ is formally divergent.
Therefore one should regularize the spacetime integrals  by using
a cut-off $\lambda$ which makes the source localized in a small,
but not exactly point-like, spacetime region and we shall  consider the
limit $\lambda \rightarrow \infty$  in those matrix elements
 where $\lambda$ appears. However we will see that in our case the matrix elements, we are interested
in, do not depend from the  regularization of the integrals.

Using Eqs.~(\ref{sean} ) and (\ref{sean2}) in Eq.~(\ref{afrikka})
the detection probabilities to the field generated  by the source  can then be evaluated with no
kind of approximation and becomes
\begin{equation}\label{ligga}
{P_{UDD}}=c^{2}_{1}|m_{eg}|^{2}\int_{t_i}^{t_f}\int_{t_i}^{t_f}dt
'dt '' e^{i\omega_{eg}(t ''-t ')} \langle
 t_{i}|\Phi(\textbf{x},t')\Phi(\textbf{x},t'')|t_{i}\rangle
\end{equation}
where we have assumed that the UDD-field interaction occurs in the
interval time $[t_f,t_i]$. Three different physical situations can
occur for: $y_{0}< t_i$, $t_i<y_0<t_f$, and $t_f<y_0$.

$i)$ $\mathbf{y_{0}< t_i}.$ The  classical point-like source is turned
on
  at $y_{0}< t_i$. In this case the response of the UDD takes the form
 \begin{eqnarray} \label{triva}
 {P_{UDD}}&=& P_1(t_f,t_i) +g^2 |P_2(s_f,s_i)|^2 \nonumber \\
& =& c^{2}_{1}|m_{eg}|^{2}\int_{t_i}^{t_f}\int_{t_i}^{t_f} dt'
dt''e^{i\omega_{eg}(t''-t')}\langle
0 |\Phi(\textbf{x},t')\Phi(\textbf{x},t'')|0 \rangle \nonumber \\
& &
+g^{2}c^{2}_{1}|m_{eg}|^{2}\Theta^2(t_i-y_0)\left|\int_{t_i}^{t_f}
dt''e^{i\omega_{eg}t''}
 \Delta(\textbf{x} -\textbf{y},t''-y_{0}) \right|^{2}
\end{eqnarray}
where  $\Delta$ is the propagator function coming from the field
commutator   $[\Phi(x),\Phi(y)] = i\Delta(x - y)$ and is vanishing
when its argument is space-like. Therefore to the last time
integral of Eq.~(\ref{triva})
 contribute only the values of  $\Delta (x)$  such that $x$ is inside the light-cone centered
on the spacetime point $y$, where the classical source is
localized. The detection  probability for UDD can be seen to made of two terms. The
first representing the  vacuum
contribution to the detector response function, is source independent and   and presents non
zero contributions  outside the light-cone  centered on the source.
The second is source dependent  and, as  shown in Appendix \ref{sec:cel2},
 $P_2(s_f,s_i)$ can be put in the form:
\begin{eqnarray}\label{passaem}
&&P_2(s_f,s_i) \nonumber
\\ & &=2c_{1}|m_{eg}|\Theta(t_i-y_{0})\Theta(s_f^{2})
 \left [\Theta (-s_i^{2})\bigg(F_{1}(0,s^2_{f})-
 \frac{e^{i\omega_{eg}(r+y_0)}}{8\pi
 r}\bigg)+\Theta (s_i^{2})F_{1}(s^2_i,s^2_{f})\right]
\end{eqnarray}
where $F_{1}(u^2,v^2)$ is defined in Eq.~(\ref{passaem2}) and
$s_{f ( i)}^2= (t_{f ( i)}-y_0)^2-r^2$ with
$r=|\textbf{x}-\textbf{y}|$. Because $ \Theta(s_f^{2})$ appears in
the expression~(\ref{passaem}), the source dependent contribution
of the UDD detector response  turns out to be automatically causally retarded

$ii)$ $\mathbf{t_i<y_0<t_f}.$ The field-classical source coupling is turned
in the time interval  $[t_f,t_i]$. We can analyze this situation
assuming that the coupling of the detector with the  field  is
 turned on from $t_i$ to $y_0 -\epsilon$ and from $y_0
+\epsilon$ to $t_f$, while the source-field interaction is
effective in the interval time
 from  $y_0 -\epsilon$ to $y_0 +\epsilon$. Then we will take the limit $ \epsilon \rightarrow 0$
in the expressions obtained. Following the same procedure used to
calculate the response of detector in the previous situation, we
obtain for the UDD probability detection
\begin{eqnarray}
{P_{UDD}}= P_1(t_f,t_i) +g^2 |P_2(s_f,0)|^2 .
\end{eqnarray}
Again $P_1(t_f,t_i)$ represents the vacuum  response contribution
and coincides with the one of Eq.~\ref{triva} while $P_2(s_f,0)$  is  linked to the variation of the source and is
given in this case by
\begin{eqnarray} \label{mussc}
P_2(s_f,0) =2c_{1}m_{eg}\Theta(s_f^{2})
 \bigg(F_{1}(0,s_{f}^2)-
 \frac{e^{i\omega_{eg}(r+y_0)}}{8\pi
 r}\bigg)
\end{eqnarray}
where we have assumed $\lim_{x\rightarrow 0_+}\Theta(x)=1$. Also
in this case we observe that the source dependent part of the UDD
 response is  vanishing outside the light-cone centered on the source.

$iii)$ $\mathbf{t_f<y_0}.$  The classical point-like source   is turned on
at $y_0>t_f$. Such  a physical situation is not of interest for
evaluating  the detector response of the field generated by the
source. In fact in this case the quantum field state describing
our system in the time interval $[t_i,t_f]$ is the vacuum state
$|0 \rangle$ and therefore the response of the UDD is simply given
by vacuum contribution with ${P_{UDD}}= P_{1}(t_f,t_i)$.

The results presented in this section show that  the use 
UDD to detect field  generated by a classical point-like source does not 
give rise to the instantaneous 
appearance of quantum correlation over the whole space
and therefore the UDD response is 
causally retarded. This causal behavior comes out naturally from
the detector  models and is not put in a sense  by ``hand''.

\subsection{Response of Glauber  detector}
In relation  to the causal response of the UDD  to the field
generated  by the classical source, localized in space and time, in our scalar QFT model it
appears also of  interest to evaluate here the response of the GD
model on order to see how realistic its use in order to to determine the structures
of quantum correlations function.  As seen in Sec.~\ref{model}, its use for free fields leads
by its very nature to the development of effects developing
over space-like distances and at variance with
Einstein's causality.

In a manner analogous to the calculation of UDD response, here we again
consider a source localized in an infinitesimal spacetime region
around the space time point$y$. The  detection probability for the Glauber
detector $P_{GD}$, assumed to be at rest and located in $\textbf{x}$, may
thus be obtained by inserting  Eqs.~(\ref{sean} ) and (\ref{sean2}) in
Eq.~(\ref{ense})
\begin{equation}\label{ligga2}
{P_{GD}}=c^{2}_{1}|m_{eg}|^{2}\int_{t_i}^{t_f}\int_{t_i}^{t_f}dt
'dt '' e^{i\omega_{eg}(t ''-t ')} \langle
 t_{i}|\Phi^-(\textbf{x},t')\Phi^+(\textbf{x},t'')|t_{i}\rangle .
\end{equation}
Again three different configurations may be considered.

$i)$ $\bf{y_0<t_i.}$ In this case the response of the GD, as shown in
the explicit calculation of Appendix, may be expressed as:
\begin{equation}\label{anguun}
{P_{GD}}= g^2 \left| \frac{1}{2}P_2(s_f,s_i) +
P_3(s_f,s_i)\right|^2
\end{equation}
with $P_2(s_f,s_i)$ and $P_3(s_f,s_i)$ given  with respect by
Eq.~(\ref{passaem}) and
\begin{eqnarray}\label{anguun3}
P_3(s_f,s_i) \nonumber
&=&c_{1}|m_{eg}|\Theta(t_i-y_{0})\bigg[\Theta (s_f^{2})
\bigg(\Theta(s_i^{2})F_{2}(s^2_i,s^2_{f})+\Theta (-s_i^{2}) \big(
F_{2}(0,s^2_{f}) +F_{3}(s^2_i,0)\big)\bigg)\nonumber \\
&+&\Theta (-s_f^{2})F_{3}(s^2_i,s^2_f)\bigg],
\end{eqnarray}
where $F_{2}(u^{2},v^2)$ and $F_{3}(u^{2},v^2)$ are defined in the
Eqs.~(\ref{passaem3}) and ~(\ref{passaem4}). We observe in
Eq.~(\ref{anguun})  that all terms are source dependent. In
particular  from  the expression~(\ref{anguun3}) we also note  the
Heaviside function $\Theta(-s^{2})$ appears  in $P_3(s_f,s_i)$.
This implies that GD may instantaneously  respond to the variation
of the source giving rise to non-locality in our model  in
agreement with what seen in Sec~\ref{model}.

$ii)$ $\bf{t_i<y_0<t_f}.$ The GD   detection probability is:
\begin{equation}\label{qwq}
{P_{GD}}= g^2 \left| \frac{1}{2}P_2(s_f,0) + P_3(s_f,r)\right|^2
\end{equation}
where $P_2(s_f,0)$ is given in Eq.~(\ref{mussc}) and $P_3(s_f,r)$
is defined as
\begin{eqnarray}
P_3(s_f,r) \nonumber =c_{1}m_{eg}\bigg[\Theta (s_f^{2})
 \bigg(F_{2}(0,s^2_{f}) +F_{3}(-r^2,0)\bigg)
+\Theta (-s_f^{2})F_{3}(-r^2,s^2_f)\bigg].
\end{eqnarray}
We note  that  again  in the  detection probability of
Eq.~(\ref{qwq}), consisting of all source dependent contributions,
non causal terms appear .

$iii)$ $\bf{t_f<y_0}$. In this case  the response of the GD vanishes,
because this detector model is not sensitive  to the vacuum fluctuations.

Finally we point   that the appearance of non-causal  terms in
the response function of GD, given  in Eqs.~(\ref{anguun}) and
(\ref{qwq}),  cannot be related to the zero-point vacuum
fluctuations, differently from what happens for UDD detector. Thus
non-causal behavior must be ascribed to the fact that the
quantity $\Phi^{-}\Phi^{+}$ which  does not satisfy the micro-causality
principle, appears  in the probability of detection for GD, given
by Eq.~(\ref{ense}), as a consequence of the standard application
of the RWA. This  is again in agreement with the previous results
showing that the use of the RWA leads to the development,over space like distances, of quantum correlations not describing
 genuine entanglement and at  variance with causal propagation of the signals
\cite{Com1,Com2,Cler}.
\section{Conclusions}\label{model3}

The evaluation of non-local quantum correlations, such as the entanglement between two or more systems consisting
of separated quanta field, can be obtained by interpreting measurements performed 
with  suitable quantum detector models not inducing  by theirselves  non-locality.
In fact  non-local effects due to the use of unappropriate model detectors 
could lead  to the development of correlations 
over the the whole space mimicking a not physical entanglement and even violating Einstein's causality. Thus the theory of detection is of  importance
in the interpretation of measurement and observability of quantum non-local effects.

Mainly two detector models are commonly used, that is GD and UDD models
\cite{Gla1,UNR,svai,sch,lag}. The difference relevant, for our
purpose, between these kinds of detector models is that in the Hamiltonian describing the GD model the
RWA is adopted and it thus  responds only to the positive frequency of the field it detects while
in the UDD model the Hamiltonian maintains the counter-rotating terms in
field-detector interaction and thus responds both to positive and
negative field frequencies.

Because a rigorous analysis of the  measurement and observability
of correlations through quantum detection processes in QFT  requires the use of
suitable models that can be solved exactly \cite{Pow}, we have used a  QFT  system,  formed by a quantum scalar
field coupled linearly to  a classical scalar source localized in
a finite spacetime region, that presenting these characteristics  can be considered a good
model\cite{Bus2}.

The use of the GD  to interpret  the appearance of quantum correlations
in  QFT models has been questioned\cite{XTTT,XTTT2,milo}.
 In the first part of this paper we have shown that by taking 
 a coherent state of a quantum scalar  field, whose the
expectation value of the field operator given by a wave plane with a
sharp front, the scalar quanta counting probability, evaluated
according to the   GD model, comes out  different from zero before the
signal reaches detector. Such a result, which is in agreement with
what  already obtained in the case of free electromagnetic field
\cite{XTTT,XTTT2}, arises also the question relative to  the role
played by RWA in the models of quantum  detection theory  and its relation
with the development of spurious quantum correlations at space-like distances.  In the same spirit here we have
extended our analysis by also examining the  detection process in the case of a quantum scalar field generated by sources.
We have then adopted a procedure, previously used for the 
electromagnetic fields case \cite{milo} that makes use of the Heinseberg picture, and then applies RWA to the formal solution of
the detector-field interaction. The fields are thus obtained from the complete
Hamiltonian, that describes the quantum field interacting both with the source
and the detector, where also the energy non-conserving terms are kept. In such models
it is possible  to calculate the detection probability rate. We have shown that by first
obtaining the fields with the complete Hamiltonian, including both conserving and 
non-conserving energy terms,  then  separating the full retarded
field in a retarded positive and negative frequency parts and finally applying the RWA on the field themselves ,  a causal rate of detection probability
is obtained. This approach, due originally to Milonni, is really different from the 
standard application of the RWA in the Hamiltonian. It does not in fact give rise to
the appearance of non-local effects in the  evolution of the positive and negative frequency parts of the field.
This deep  difference  in the final results must be associated to the fact
that the spectral decomposition of the retarded positive (negative) frequency
part of the field does contain both positive and negative frequencies coming from the $\Theta$-like retarded terms.
 This is therefore 
different from the standard procedure adopted in the Glauber theory of detection where only
positive or negative frequencies  of the complete field are kept. Thus such an application of the RWA leads to an
 effective new detector model,  which  differs from the GD one and  does not give rise
to  quantum correlations at space-like distances, but the causal behavior  results  to be
put  by ``hand'' in the solution of complete detector-field interaction.

 We have shown that in our scalar model  local operator function of the field
develop causally from the source, nevertheless  non-local effects appear in the expectation values of
one-point positive localization observables, such as the Glauber and
Newton-Wigner
 operators. The reason of this result is however that these operators do not satisfy the micro-causality principle and therefore induce,
by their very  definition,  effects over space-like distances
\cite{Bus} . Thus  they must not be used, for example by calculating their  correlations  to furnish
indications of the presence of non-local effects.

A valid interpretation of the appearance of quantum correlations in our model
requires to analyze realistic models that describe  the 
detection of the scalar field generated by the source. To this end
we  have explicitly evaluated the response function of two detector models, that is the UDD and
GD models, at rest in our reference frame in order to avoid the appearance of  Unruh like
effects. We have then shown  that the UDD detection probability causally 
responds  to the field generated  by the
source and is not characterized by non-local effects,apart from the source-independent vacuum contribution
 related to the UDD  sensitiveness  to zero point  fluctuations.
 Anyway this term must not be considered  to describe the appearance of non-locality   due
to the variations of the source and can in principle be taken into account to  interpret the results
\cite{Com1,Com2,Com3}. On the other hand  the response function of the GD
model gives   source dependent terms in the detection probability rate that correspond to an instantaneous
spreading of source effects over the whole space. If taken at face value this
 would seem to imply a violation of causal propagation of signals in our QFT
 system and therefore our results confirm  previous ones  indicating that the  standard use of
RWA  does lead to development of non-local effects with time
\cite{Com1,Com2,Cler}.

In conclusion to   measure either the   quantum correlations or those
causal effects linked to the time varying sources in our  QFT model, the adoption  of the GD model, with the RWA in the Hamiltonian
turns out to be unappropriate inducing by its very definition  non-locality. Another detector model, the MD, obtained by anon-standard
application of the RWA in the Heisenberg picture field solution, although it  guarantees  causality, presents the characteristic that its behavior is somehow  imposed in the detection theory from the outside and one may then not be sure whether source relevant terms may also be thrown out. 
Instead the UDD model must be preferred to describe quantum correlation for quantized fields because  the
appearance of non-local effects  in its response to field quanta is only due the zero
point vacuum fluctuations and does not depend from the source. Moreover this  behavior comes out naturally from the
detection model itself. It would then be also of interest to analyze the
behavior of such a quantum detector model when  it or the source
is  in arbitrary motion in order to study the relation between the
appearance of Unruh effects and non-local quantum effects.

\appendix
\section{Photodetection probabilities}\label{sec:cel2}
Here we shall  give the explicit calculation of the detection
probability of the UDD and GD  in our QFT system when  $y_0<t_i$.

For the UDD the source dependent contribution of response to the
field is given by the second term of the right side  of the
expression~(\ref{triva}). The integral appearing in it,  after
inserting the explicit form of $\Delta$ function,  can be
expressed as
\begin{eqnarray} \label{anguun2}
\lefteqn{ \int_{t_i}^{t_f}  d
t''e^{i\omega_{eg}t''}
 \Delta(\textbf{x} -\textbf{y},t''-y_{0})    {}}\nonumber \\
& & =2 e^{i\omega_{eg}y_{0}}\int_{y_{0}}^{t} d
t''e^{i\omega_{eg}(t''-y_{0})}
\Bigg[\!\!-\!\frac{1}{4\pi}\delta(s''^{2})+\!\frac{m\,\Theta(s''^{2})}{8\pi\sqrt{s''^{2}}}J_{1}(m\sqrt{s''^{2}})\!\Bigg]
\end{eqnarray}
 with $s''^{2}=(t''-y_{0})^{2}-r^{2}$ and     $J_{1}$ indicating    the Bessel function of first order \cite{Weiss}.
Performing the change $t''\rightarrow s''^{2}$ in integration
variable ($s^{2}(t)$ is monotone in the integration variable for
$t''$) Eq.~(\ref{anguun2}) may be put in the form:
\begin{eqnarray}
& & \int_{t_i}^{t_f}  d t''e^{i\omega_{eg}t''}
 \Delta(\textbf{x} -\textbf{y},t''-y_{0})\nonumber \\
& =&\Theta(s_f^{2})
 \Bigg \{ \Theta (-s_i^{2})\Bigg[- \frac{e^{i\omega_{eg}(r+y_0)}}{8\pi
 r}+e^{i\omega_{eg}y_0}\int_{0}^{s_f^{2}}\frac{ d
s''^{2}}{2\sqrt{s''^{2}+r^{2}}}\,e^{i\omega_{eg}\sqrt{s''^{2}+r^{2}}}
\,\frac{m}{8\pi\sqrt{s''^{2}}}J_{1}(m\sqrt{s''^{2}})\Bigg] \nonumber \\
&+&\Theta (s_i^{2})\int_{s_i^2}^{s_f^{2}}\frac{ d
s''^{2}}{2\sqrt{s''^{2}+r^{2}}}\,e^{i\omega_{eg}\sqrt{s''^{2}+r^{2}}}
\,\frac{m}{8\pi\sqrt{s''^{2}}}J_{1}(m\sqrt{s''^{2}})\Bigg\}
\end{eqnarray}
By using the last equation   and    defining the function $
F_{1}(u^2,v^2)$ as:
\begin{equation} \label{passaem2}
 F_{1}(u^2,v^2) = 2e^{i\omega_{eg}y_{0}}\Bigg\{e^{i\omega_{eg}r}\left(-\frac{1}{8\pi
r}\right)\!+\!\int^{u^2}_{v^2}\!\!\!\frac{ d
s''^{2}}{2\sqrt{s''^{2}+r^{2}}}\,e^{i\omega_{eg}\sqrt{s''^{2}+r^{2}}}
\,\frac{m}{8\pi\sqrt{s''^{2}}}J_{1}(m\sqrt{s''^{2}})\Bigg\},
\end{equation}
we  obtain  the expression~(\ref{passaem}), whose square modulus
gives   the  source dependent term of UDD detection probability in
our  QFT model.

Now we evaluate explicitly the response function of the GD  when
$y_0<t_i$. To   this purpose we insert Eq.~(\ref{sean}) in
(\ref{ligga2}) and we obtain:
\begin{equation}\label{lunko}
P_{G}(t_f,t_i)=c^{2}_{1}|m_{eg}|^{2}g^{2}\Theta^{2}(t-y_{0})
\Bigg|\int_{t_{i}}^{t_f} d t''e^{i\omega_{eg}t''}
\Delta_{+}(\textbf{x}-\textbf{y},t''-y_{0})\Bigg|^{2}.
\end{equation}
The square modulus in the above expression can be written after
decomposing $\Delta_{+}$ in its real and imaginary parts as:
\begin{eqnarray}\label{respo2}
 \Bigg| \int_{t_i}^{t_f}  d t''e^{i\omega_{eg}t''}
\Delta_{+}(\textbf{x}-\textbf{y},t''-y_{0})\Bigg|^{2}  =
 \Bigg| \int_{t_i}^{t_f} d t''e^{i\omega_{eg}t''}
\textrm{Re}\Delta_{+} +i\int_{t_i}^{t_f} d
t''e^{i\omega_{eg}t''} \textrm{Im}\Delta_{+}\Bigg|^{2}
\end{eqnarray}
Because  $\textrm{Re}\Delta_{+}=\frac{1}{2}\Delta$ it can be easily shown
that  the first term in  (\ref{respo2}) gives a contribution
proportional to    the expression~(\ref{anguun2})
\begin{equation}\label{stink}
\int_{t_i}^{t_f} d t''e^{i\omega_{eg}t''}
\textrm{Re}\Delta_{+}(\textbf{x}-\textbf{y},t''-y_{0})=\frac{1}{2} \int_{t_i}^{t_f} d t''e^{i\omega_{eg}t''}
\Delta(\textbf{x}-\textbf{y},t''-y_{0})
\end{equation}
Instead using  the explicit forms of
$\textrm{Im}\Delta_{+}(x)$ for  $x_{0}>0$ \cite{Weiss}
\begin{eqnarray}
 \textrm{Im}\Delta_{+}(x)&=&-i\Big[\frac{m\Theta(x^{2})}{8\pi\sqrt{x^{2}}}N_{1}(m\sqrt{x^{2}})+\frac{2m\Theta(-x^{2})}{8\pi^{2}\sqrt{-x^{2}}}K_{1}(m\sqrt{-x^{2}})\Big]
\end{eqnarray}
with  $N_{1}(z)$ and $K_{1}(z)$ respectively the first order
 Neumann and  Mac Donald functions and performing the change in the integration variable  $t''\rightarrow
s''^{2}$ we can evaluate the second integral in
 the right side of (\ref{respo2}):
\begin{eqnarray}\label{nongaucau}
 & & i\int_{t_1}^{t_f} d t''e^{i\omega_{eg}t''}
\textrm{Im}\Delta_{+}(\textbf{x}-\textbf{y},t''-y_{0})=
  \nonumber  \\
  & & \Theta (s_f^{2})\Bigg[\Theta(s_i^{2})e^{i\omega_{eg}y_{0}}\int_{s_i^2}^{s_f^{2}}\frac{ d
s''^{2}}{2\sqrt{s''^{2}+r^{2}}}\,e^{i\omega_{eg}\sqrt{s''^{2}+r^{2}}}
\,\frac{m}{8\pi\sqrt{s''^{2}}}N_{1}(m\sqrt{s''^{2}})\nonumber \\ & &+
\Theta (-s_i^{2}) \bigg(e^{i\omega_{eg}y_{0}}\int_{0}^{s_f^{2}}\frac{ d
s''^{2}}{2\sqrt{s''^{2}+r^{2}}}\,e^{i\omega_{eg}\sqrt{s''^{2}+r^{2}}}
\,\frac{m}{8\pi\sqrt{s''^{2}}}N_{1}(m\sqrt{s''^{2}})\nonumber \\ & &
+e^{i\omega_{eg}y_{0}}\!\!\int_{s_i^{2}}^{0}\frac{ d
s''^{2}}{2\sqrt{s''^{2}+r^{2}}}\,e^{i\omega_{eg}\sqrt{s''^{2}+r^{2}}}
\,\frac{2m}{8\pi^{2}\sqrt{-s''^{2}}}K_{1}(m\sqrt{-s''^{2}}) \bigg)\Bigg] \nonumber \\  & &
+\Theta (-s_f^{2})e^{i\omega_{eg}y_{0}}\!\!\int_{s_i^{2}}^{s_f^{2}}\frac{ d
s''^{2}}{2\sqrt{s''^{2}+r^{2}}}\,e^{i\omega_{eg}\sqrt{s''^{2}+r^{2}}}
\,\frac{2m}{8\pi^{2}\sqrt{-s''^{2}}}K_{1}(m\sqrt{-s''^{2}})
\end{eqnarray}

Now inserting  the Eqs.~(\ref{stink}) and (\ref{nongaucau}) in
(\ref{lunko}) and defining the functions $F_2(u^2,v^2)$ and
$F_3(u^2,v^2)$ as
\begin{equation} \label{passaem3}
 F_{2}(u^2,v^2) = e^{i\omega_{eg}y_{0}}\int_{v^2}^{u^2}\frac{ d
s''^{2}}{2\sqrt{s''^{2}+r^{2}}}\,e^{i\omega_{eg}\sqrt{s''^{2}+r^{2}}}
\,\frac{m}{8\pi\sqrt{s''^{2}}}N_{1}(m\sqrt{s''^{2}})
\end{equation}
\begin{equation} \label{passaem4}
 F_{3}(u^2,v^2) = e^{i\omega_{eg}y_{0}}\!\!\int_{v^2}^{u^2}\frac{ d
s''^{2}}{2\sqrt{s''^{2}+r^{2}}}\,e^{i\omega_{eg}\sqrt{s''^{2}+r^{2}}}
\,\frac{2m}{8\pi^{2}\sqrt{-s''^{2}}}K_{1}(m\sqrt{-s''^{2}}),
\end{equation}
with respect, we obtain for the GD response the
expression~(\ref{anguun}).

\end{document}